\documentclass[12pt,preprint]{aastex}

\usepackage{mathptm}
\usepackage{graphicx}
\usepackage{url}

\begin{document}

\title{Observation of 2011-02-15 X2.2 flare in Hard X-ray and Microwave}

\author{Natsuha Kuroda$^1$, Haimin Wang$^1$, Dale E. Gary$^1$}

\affil{1. Center for Solar-Terrestrial Research, New Jersey Institute of Technology, University Heights, Newark, NJ 07102-1982, USA}

\email{nk257@njit.edu}

\begin{abstract}
Previous studies have shown that the energy release mechanism of some solar flares follow the Standard magnetic-reconnection model, but the detailed properties of high-energy electrons produced in the flare are still not well understood. We conducted a unique, multi-wavelength study that discloses the spatial, temporal and energy distributions of the accelerated electrons in the X2.2 solar flare on 2011, Feb. 15. We studied the source locations of seven distinct temporal peaks observed in hard X-ray (HXR) and microwave (MW) lightcurves using the Reuven Ramaty High Energy Solar Spectroscopic Imager (RHESSI) in 50 -- 75 keV channels and Nobeyama Radioheliograph (NoRH) in 34 GHz, respectively. We found that the seven emission peaks did not come from seven spatially distinct sites in HXR and MW, but rather in HXR we observed a sudden change in location only between the second and the third peak, with the same pattern occurring, but evolving more slowly in MW. Comparison between the HXR lightcurve and the temporal variations in intensity in the two MW source kernels also confirmed that the seven peaks came predominantly from two sources, each with multiple temporal peaks. In addition, we studied the polarization properties of MW sources, and time delay between HXR and MW. We discuss our results in the context of the tether-cutting model.

\end{abstract}


\section{Introduction}

In the study of solar flares, a widely accepted model of the physical mechanism of flares is the Standard magnetic reconnection, referred to as the Carmichael, Sturrock, Hirayama, Kopp and Pneuman, or "CSHKP" model, which assumes magnetic reconnection as the principle mechanism responsible for accelerating flare particles (Hudson 2011). In this model, magnetic reconnection occurs somewhere above the magnetic loop structure of the active region. The form of this magnetic reconnection mechanism is currently not fully understood, although hypotheses such as a large-scale or a collection of small-scale DC electric fields (Dreicer fields) have been suggested (Litvinenko et al. 1996). Magnetic reconnection gives considerable energy to the particles that are present at the reconnection site, in some cases accelerating them to the relativistic regime.

The dynamics of these accelerated particles after leaving the reconnection site is still not well understood. However, one key observation is that, among the many wavelengths at which flare energy is emitted, hard X-ray (HXR) and microwave (MW) emissions show time profiles that are very well-correlated during the impulsive phase of the flare (e.g. Kundu et al. 1982, Cornell et al. 1984, Asai et al. 2012). This observational fact has led many to the conclusion that the same population of electrons is responsible for producing HXR and MW emission, and many studies have been done on various aspects of this behavior. Kundu et al. (1982) analyzed the dynamics of 1980 Jun 25 flare using MW time profile, MW images, and HXR time profile. Cornell et al. (1984) investigated the causes of the time delay between corresponding HXR and MW peaks, and Asai et al. (2012) studied several flares that had corresponding HXR and MW time profiles in terms of electron spectral indices. 

One of the interesting topics related to this correlation, and which will be discussed in this paper, is the delay between their peak times. It has been observed that MW peaks often lag HXR peaks. The time scale of this delay varies from subsecond to sometimes more than ten seconds. Cornell et al. (1982) discussed that the delay observed in their study (an average of 0.2 s) can be explained by the existence of two components of energetic electrons: a prompt component that immediately escapes from the acceleration region, and a delayed component that first undergoes pitch-angle scattering and thus is less impulsive. They discussed that the first component would be the population which produces HXR and the second component would be the population which produces MW, and therefore MW peaks lag HXR peaks. Gary \& Tang (1985) studied a single-peaked event in HXR and MW, and found that the delay they found can be explained simply by the difference in decay time of HXR and MW emission (Gary \& Tang 1985). They claimed that the two emission processes share the same acceleration profile, but not necessarily the same population; MW-emitting electrons are those that are trapped within the magnetic fields (because of magnetic mirroring), where the HXR-emitting electrons are those that directly hit the chromosphere. The MW emission has longer decay time than HXR emission because of prolonged trapping, which accounts for the observed delay in the MW peak. Silva et al. (2000) conducted a statistical study of 57 simultaneous peaks of HXR and MW, and found that most probable delay time to be 2 -- 4 seconds, and that the delay time decreased as the HXR energy increased, which supports the hypothesis that the MW-emitting electrons are the higher-energy counterpart of the HXR-emitting electrons. However, they also found that the electron energy spectral index for MW is harder than that for HXR by 0.5 -- 2.0 on average for 75\% of the bursts, which was interpreted as due to an upward break in the energy spectrum of the accelerated electrons responsible for HXR and MW emission. They give three hypotheses that could account for their results, (1) HXR- and MW-emitting electrons are different populations that are accelerated by different means at different sites, (2) MW-emitting electrons are accelerated by second-step acceleration, or (3) accelerated electrons follow so-called "trap-plus-precipitation" model, which explains that some of the accelerated particles that get trapped, and give off MW emission, eventually "precipitate out" due to Coulomb collisions, and become those that hit the chromosphere and give off delayed HXR emission (Melrose et al. 1976). This same model was used by Aschwanden (1995) to study two different components of HXR emission.

In terms of the relative locations of HXR and MW emission, HXR emission can come from loop-top (above-the-loop) and/or footpoint sources, and for MW it can come from either a combination of loop-top and loop-legs or from footpoint sources, depending on the MW frequency. For HXR energies above $\sim$ 12 keV, the emission comes from loop-top and/or footpoint sources (Guo et al. 2012, Jeffrey \& Kontar 2013, Masuda et al. 1994, Ishikawa et al. 2011, Torre et al. 2012). Loop-top sources observed at this energy range have been considered as evidence for the particle acceleration region. They can sometimes be observed simultaneously with footpoint sources (Ishikawa et al. 2011, Kuznetsov et al. 2014). For example, Kuznetsov et al. (2014) analyzed the spatial properties of the 2004, May 21 flare, and found a loop-top source in 12 -- 25 keV range, a loop-top source with footpoint sources in 25 -- 50 keV range, and only footpoint sources in 50 -- 100 keV range. Above $\sim$ 50 keV, footpoint sources start to clearly dominate. 
For MW in the range 3 GHz $\sim$ 30 GHz, the dominant emission mechanism in a flare is gyrosynchrotron radiation produced by electrons with mildly relativistic energy distribution. A schematic model of such emission (Bastian et al. 1998, Gary et al. 2013) reveals that, at lower frequencies (optically thick regime), the magnetically weak, loop-top source has higher brightness temperature than the magnetically stronger footpoint sources. However, because the loop-top source has a lower peak frequency than footpoint sources and the brightness temperature of the loop-top source falls off more steeply than that of footpoint sources in the optically thin regime, the footpoint sources have higher brightness temperature than the loop-top source at higher frequencies. In case of an asymmetric loop, the magnetically weaker footpoint reaches its optically thin regime at lower frequency than the magnetically stronger footpoint does, thus the magnetically strong footpoint dominates at higher frequencies. Such is indeed observed: the MW source variously appears as a loop-top source or a whole-loop structure with enhanced loop-top (Asai et al. 2013, Kushwaha et al. 2014, Kuznetsov et al. 2014) at low frequencies, as well as single or double footpoint sources at higher frequencies (Kundu et al. 1982, Asai et al. 2013). MW footpoint sources usually match or tend to be spatially very close to HXR footpoint sources in near-disk-center flares. However, this is not the case for near-limb flares, since MW and HXR emissions are produced at different heights (MW emission is produced at the magnetic mirroring point, which is above chromosphere). Also, Sakao et al. found that, in double footpoint configuration with asymmetrical magnetic strength, stronger HXR emission tends to originate from the weaker magnetic footpoint (Sakao et al. 1996). They argue that the magnetically stronger side of the loop has stronger magnetic mirroring, thus allowing fewer energetic particles to penetrate, producing weaker HXR emission. Since stronger MW footpoint emission comes from magnetically stronger footpoint as mentioned before, this may result in the discrepancy between HXR and MW emission source location in some cases.

In this study, we analyzed the impulsive phase of the X2.2 solar flare that occurred on 2011, February 15, which showed correlated HXR and MW time profiles. This event has received considerable attention in the literature (Zharkov et al. 2013, Kerr et al. 2014, Milligan et al. 2014, Inoue et al. 2014, Wang et al. 2014) because it was the first X-class flare of the current Solar cycle and was successfully covered by many different instruments. Summarizing these studies, Zharkov et al. (2012) analyzed the seismic/sunquake response, Kerr et al. (2014) analyzed the white-light emission, Milligan et al. (2014) calculated the total energy radiated by the lower solar atmosphere (optical, ultra-violet, and extreme ultra-violet wavelength) and that of the non-thermal electrons derived from HXR. Wang et al. (2014) analyzed flow motion and sunspot rotation of the active region during the impulsive phase. Inoue et al. (2014) performed a magnetohydrodynamic simulation based on non-linear force-free field for the active region responsible for this flare. This paper focuses on the study of flare accelerated electrons, and we do so by observing the flare in HXR and MW in high-temporal (sub-second) and spatial (one to few arcsec) resolutions. To our knowledge, this is the first time the high-resolution comparison between HXR and MW emission of this flare was drawn. We observed that this flare consisted of several temporally distinct peaks in HXR and MW, and we spatially resolved them to see if they were coming from the same or different sources.

In section 2 we list the sources of data we used for the study.  In section 3 we describe the analysis in terms of energy/frequency-dependent time profiles and images.  We give the results of the analysis in section 4, and conclude in section 5.

\section{Data}

The X2.2 flare occurred on Feb. 15, 2011, starting from approximately 0130~UT and ending at about 0700~UT, from the active region 11158 located at about S21W28. We used data from the Reuven Ramaty High Energy Solar Spectroscopic Imager (RHESSI; Lin et al. 2002) for HXR observation and Nobeyama Radio Observatory (NRO; Nakajima et al. 1985, Nakajima et al. 1994) for MW observation. We also used data from the Atmospheric Imaging Assembly (AIA; Lemen et al. 2012) on board the Solar Dynamic Observatory (SDO) for extreme ultraviolet observation, and magnetogram data obtained from the Helioseismic and Magnetic Imager (HMI) on board the SDO.

RHESSI is capable of imaging solar flares in the energy ranges from soft X-rays ($\sim$ few keV) to gamma rays (up to $\sim$ 20 MeV). Its default temporal cadence is 2 seconds, and the spatial cadence can go down to 1 arcsec/pixel depending on energy range. The temporal cadence can be reduced down to 0.1 second, if one uses demodulation code hsi\_demodulator.pro of the SSW software (Arzner 2004, Qiu et al. 2012).

For MW observation, we used data from both the Nobeyama Radio Polarimeters (NoRP) and Nobeyama Radioheliograph (NoRH) available from NRO. NoRP records the total incoming flux from the Sun in 1, 2, 3.75, 9.4, 17, 35, and 80 GHz. NoRH produces the full-disk images of the Sun in 17 and 34 GHz with a cell size of 10 arcsec/pixel and 5 arcsec/pixel, respectively, and a spatial resolution rougly 2 times larger.  The instrument uses the solar disk for determining absolute position and brightness scale, assuming that the solar disk has a uniform brightness temperature of $10^4$~K at 17~GHz.  However, during bright solar flares such as this one, the solar disk may be relatively too weak for a precise determination of absolute position, and the source positions can display a jitter of roughly 5~arcsec. We discuss this further in section 4.2. The temporal cadence we used is 0.1 second and 1 second for NoRP and NoRH, respectively.

The AIA on board the SDO observes the full-disk solar atmosphere in 10 different energy channels from white-light continuum (photosphere) to Fe VIII line (coronal flaring regions). Its temporal cadence is 10 to 12 seconds and the spatial resolution is 0.5 to 1 arcsec/pixel. The HMI on board the SDO measures the line-of-sight component of the photospheric magnetic field in two intervals; 45 seconds (with the Doppler camera) and 720 seconds (computed from the vector field camera data). The spatial resolution is 1 arcsec/pixel.

\section{Analysis}

To investigate the properties of flare accelerated electrons, we analyzed the temporal and spatial profiles of the impulsive phase of the flare using instruments listed above. As will be shown, the impulsive phase of this flare (about 10 minutes) consisted of several temporally distinct peaks in HXR and MW. Each peak, according to the standard magnetic reconnection model, comes from a temporally distinct magnetic reconnection event which produces highly accelerated electrons. The aim of this study is to spatially resolve these temporal peaks, in both HXR and MW, to find if they were coming from different or the same source locations. We also compare their locations against extreme ultraviolet and magnetogram images. We investigate time delays between HXR and MW peaks, and examine their relationship to the spatial and temporal analysis results.

\subsection{HXR and MW time profile}

RHESSI successfully covered the entire period of the impulsive phase of this flare, which is from about 0148UT to 0158UT.  We plot the HXR lightcurve during this period in 50 -- 75 keV energy channels. The time profile is shown as a black curve in Figure 1. In order to accomplish the highest temporal resolution, we utilize demodulating code as described in Qiu et al. (2012), and reduced its default cadence to 0.2 -- 0.3 seconds.

The NoRP and NoRH successfully observed the entire period of this flare. The time profiles in these frequencies are plotted in Figure~2. In this flare, the MW time profiles for 9.4~GHz and 17~GHz showed the best correlation with the 50 -- 75 keV HXR time profile. The time profiles for MW at 9.4~GHz and 17~GHz during impulsive phase are shown as the red and blue curve, respectively, in Figure 1. Note that 17 GHz flux is artificially increased by 500~SFU, to be on the same plotting window as the 9.4~GHz curve.

We chose HXR peaks that were sufficiently distinct and had correlating MW peaks, and calculated the peak times. Those peaks are marked with short grey, red, and blue lines for HXR, MW 9.4 GHz, and MW 17 GHz, respectively, in Figure 1. To calculate the peak time, we fitted a third-degree polynomial over a certain data range that includes a visually distinguishable spike in the target peak, and chose several other ranges to obtain several peak times, from which their median was taken as the peak time and their standard deviation was taken as the peak time uncertainties. The peak times and their uncertainties for each frequency are summarized in Table 1.

\subsection{HXR and MW imaging}
\label{subsect:HXR imaging}

We next analyzed the spatial locations of these peak emissions. For HXR, we utilized RHESSI OSPEX software prepared by the RHESSI team. We used the PIXON algorithm in the 50 -- 75 keV energy range, with 1 arcsec/pixel resolution, using the front segments of detectors 1 and 3 -- 8 with 7 -- 36 seconds integration time. We determined appropriate integration times for each peak based on the shape of the lightcurve; we chose intervals during which lightcurves showed single and clear rise-and-fall behaviors. We also identified MW emission sources, obtaining images from NoRH in 17 GHz and 34 GHz, in stokes I, at peak times in the 17 GHz time profile. The images in 34 GHz were synthesized by the readily available Hanaoka program (Hanaoka et al. 1994) from the Nobeyama Observatory. We overplotted HXR and MW energy contours for each peak separately on the HMI continuum image at 01:47UT, and they are shown in Figures 3 and 4. We also overplotted HXR and MW 34 GHz contours on the EUV images from AIA HeII 304 channel, which usually show flare ribbons, and they are shown in Figure 5. Note that AIA images for peak 3 and 4, and peak 5 and 6 are identical because we chose minimally saturated images at the time closest to HXR peak time, and images between these peaks were much more saturated.

\section{Observational results}

\subsection{HXR emission sources}

In HXR, most emission sources were identified to be footpoint sources. The energy range that we used, 50 -- 75 keV, is well within the non-thermal electron energy range that is known to show footpoint sources in the thick-target model (Saint-Hilaire et al. 2010). In Figure 3, they can all be seen as double sources over the polarity inversion line as well. Their locations all coincide with flare ribbons seen in AIA He 304 in Figure 5, which further supports that they are the footpoints of magnetic loops. For peak 1, the type of the source is uncertain. Since it is a single source, it could be a loop-top source or a single footpoint source. For peak 2, there is a single elongated source to the west of double-footpoint sources. We confirmed that this is the same source as the single remote western source observed at peak 3. The single remote source in peak 3 is one of the footpoint sources expected in a quadrupolar loop configuration, and this is also observed by Wang et al. (2012). Interestingly, the location of this source coincides with one of the seismic/sunquake sources observed by Zharkov et al. (2013). For peak 4, the somewhat elongated shape of the source was likely caused by the reappearance of western footpoint source observed at peak 2, since the elogation is directed along the flare ribbon, shown in Figure 5. For the rather spatially complex peak 7, analysis with shorter integration time revealed that the weakest, north-western source appeared only at the end of the integration time of peak 7, and therefore most of the emissions come from the pair of north-eastern and southern footpoint sources.

It is clear from the observation of the entire period that there are two distinct double-footpoint sources involved; the one best exemplified by the HXR image at peak 2, and another one responsible for the emission during the later peaks, 3 to 7. This suggests that there are at least two distinct magnetic reconnection sites for this flare. In order to make our results more quantitative, we conducted the following analysis. First, we identified source kernel pixels by taking image pixels that are above 80 percent of the maximum intensity of the image at peaks 2 and 3. The image of these kernel pixels are shown in Figure 6. We produced a running average movie of HXR images in 50 -- 75 keV with 4 seconds between and 40 seconds over each frame, and used this movie to plot the time profiles of the mean value of identified kernel pixels (movie provided as the online supplemental material). The result is shown in Figure 7. The black curve indicates the "total intensity" HXR curve, which is the same curve as the black curve in Figure 1, but with a 4~s resolution. The red curve is the time profile of the mean kernel intensity of the source observed at peak 2 (western source), and the green curve is the time profile of the mean kernel intensity of the source observed at peak 3 (eastern source). The red and green vertical line marks the peak time for peak 2 and 3, respectively. The plot clearly shows the aforementioned change of double-footpoint sources; the sharp rise in intensity of the western source and the low intensity of the eastern source at peak 2 confirm that the emission was purely from the western source for peak 2. The rest of the peaks, peak 3 to 7, come mainly from the eastern source. 

Finding two spatially distinct HXR sources, we attempted to investigate if they have different spectral properties by conducting imaging spectroscopy analysis, obtaining images from 30 keV to 100 keV with 5 keV interval. However, both sources become unrecognizable above 50 keV, preventing us from calculating meaningful spectral indices for them. 

\subsection{MW emission sources}

For MW, Figure~4 shows the 17~GHz and 34~GHz images at the peak times of the 17~GHz time profile. As mentioned in the introduction, MW sources can be either a combination of loop-top and loop-legs or footpoint sources, depending on the MW frequency. Since this flare shows sources that are spatially close in both frequencies, we assume that the source type is the same for both frequencies. We first analyze the case of footpoint sources. We know that in this flare, the peak frequency of MW emission during the impulsive phase is between 9.4 GHz and 17 GHz or higher based on the NoRP total power data (see Figure 2), so the 17 GHz and 34 GHz emissions are clearly in the optically thin regime. If the observed sources are footpoint sources, then this  fits the scenario for the schematic model mentioned in the introduction; the emission is produced by gyrosynchrotron radiation from electrons with non-thermal energy distribution, which is dominated by footpoint sources as one reaches to optically-thin frequency range. However, in this scenario the MW sources should appear as double-footpoint sources straddling the neutral line, whereas, in Figure 4, the MW sources in peaks 1 and 2 are single sources, and in later peaks, when the MW sources are double sources, the pair are oriented along and well south of the neutral line.  Because of the limited NoRH spatial resolution, together with the positional uncertainty mentioned in Section 2, we consider it likely that the sources in all cases should be shifted north by at least 5-10~arcsec (although we did not do so in Figure~4), and are in reality filled-loop or unresolved double-footpoint sources oriented across the neutral line. We also note that the projection effect may be taking place as well, since the active region is located in the southern hemisphere and the MW emission should be generated at some height above the photosphere even for footpoint sources (as mentioned in Introduction); such a projection would shift the sources south relative to the surface. Support for this interpretation is provided by the NoRH circular polarization maps shown in Figure~8 (available only at 17~GHz, since NoRH measures the 34~GHz emission in total intensity only).  Hanaoka (1996) and Nishio et al. (1997) showed that, if there are two unresolved MW sources with opposite polarities, they can appear as a single source with a gradient of the degree of polarization. Such is indeed observed in our case, where Figure~8 shows that peaks 3--7 to all have a N-S gradient in degree of polarization. In contrast, peak 2 has mainly E-W-oriented gradient, but from a location where the orientation of the neutral line turns more N-S. Peak 1 source may be different from all other sources, like in the case of HXR discussed in previous section. This therefore suggests that, at least for peak 2--7, our MW sources could be optically-thin, unresolved double-footpoint sources.  Moreover, Figure~8 shows that the source with left-circular polarization (north side) has higher degree of polarization, suggesting that the magnetic field is stronger on the negative side of the polarity inversion line as observed. At the same time, Figure~3 shows that the HXR sources are typically stronger on the opposite, positive polarity side of the polarity inversion line, so that this interpretation also agrees with the findings of Sakao et al. (1996), who found that HXR and MW sources in loops of asymmetric magnetic strength are strongest on opposite sides of the polarity inversion line.

Although the above discussion mentions the idea that each MW source is unresolved double-footpoint sources, recent NoRH observations of larger loops (Reznikova et al. 2009; Reznikova et al. 2010), where the morphology evolution is more easily seen, indicate that the MW loops tend to start out dominated by footpoint sources but then quickly evolve to loop-leg or even loop-top dominated sources, and this is likely the case also in our event.  This is the result of the accumulation of mildly relativistic electrons due to trapping in the loop, and can also explain why the sudden jump in location from western to eastern source, seen in HXR between peaks 2 and 3, is delayed and evolves more slowly in MW.  To investigate this more quantitatively, we repeated the kernel light-curve analysis done for HXR, in Figures~6 and 7, for MW. It is clear from Figure~4 that there are three distinct MW emission sources; the one for peak 1, another one for peak 2, and the one that appears at peak 3 and continues to dominate for the rest of the peaks. Comparing to the HXR sources that were discussed in previous section, it is striking to note the similarity in which the sources change.

To identify the source kernel pixels, we chose 34~GHz images and used intensity thresholds of 90 percent for peak 2 source and 60 percent for peak 3 (we decreased the threshold for peak 3 because it was weaker than the peak 2 source at that time and we could not collect kernel pixels when setting the threshold above 70 percent of the maximum intensity). The kernel pixel locations are shown in Figure 9. We plotted time profiles for the two sources against the spatially integrated 34 GHz NoRH signal. The result is shown in Figure 10, with the same color scheme as for the HXR curves in Figure 7. The red curve is the time profile for the western source, and the green curve is the time profile for the eastern source. The red and green vertical lines mark the peak time for peaks 2 and 3, respectively. Just as in HXR, it is seen that the peak 2 emission was purely from western source, and the eastern source first appears at peak 3.  However, peak 3 does not become dominant in MW until a time between peaks 3 and 4, due to the longevity of trapped particles in the western source.
By the time of peak 4, it is clear 
that the eastern source was the only source responsible for all peaks. In summary, we observed the exact same source change behavior in MW as was seen in HXR. It is also clear from Figure 5 that MW sources are all located very close to HXR sources. Therefore, the observations show evidence that there are two different magnetic reconnection sites---one responsible for emission in both HXR and MW at peak 2, and another for peaks 3 to 7.

We note that the peak brightness temperatures at 17~GHz in this event are 40-70~MK, and, since the emission is optically thin, the kinetic temperature must exceed 100~MK, so the emission is undoubtedly due to nonthermal electrons. We therefore calculate the electron energy spectral index ($\delta$) using the equation $\alpha$ = log(F$_{35}$/F$_{17}$)/log(35 GHz/17 GHz), where F$_{35}$ and F$_{17}$ indicates the flux at 35 GHz and 17 GHz respectively, and the approximation for gyrosynchrotron emission from nonthermal electron energy distribution derived by Dulk (1985), $\delta$ = (1.22 - $\alpha$)/0.9. The result is shown in Figure 11. We also show $\delta$ derived from HXR, which were calculated by Milligan et al. (2014), assuming a collisional thick-target model with a power-law electron spectrum at higher energies (Milligan et al. 2014). Blue vertical lines indicate HXR peak times. We see that the $\delta$ for HXR is around 5 $\sim$ 7, and $\delta$ for MW is around 2 $\sim$ 3. The difference between them is about 3 $\sim$ 4, which is much larger than the result found by the statistical study of Silva et al. (2000) (0.5 -- 2). Their study also shows that the mean value of $\delta$ for HXR is 5.8 $\pm$ 0.8, and that for MW is 4.8 $\pm$ 1.0. Therefore, this seems to indicate that the MW indices we found may be harder than usual by 1 $\sim$ 2 in this scenario.

\subsection{Time delays between HXR peaks and MW peaks}
As mentioned in the Introduction, in addition to the correlation between MW and HXR time profiles, it has been observed that MW peaks sometimes lag HXR peaks. We conducted an analysis on this aspect as well. Table 1 summarizes the delay time of MW 9.4 GHz and 17 GHz peaks respect to the corresponding HXR peaks. Note that, for peaks 5 and 6, we conclude that there are virtually no delays between HXR and either MW frequency, since delays are smaller than 2$\sigma$. We would also omit analysis for peak 2 since HXR signal for this peak involves two visually distinguishable spikes, which poses a problem in our method of determining the peak time and its uncertainty (see section 3.1).

Summarizing other peaks (1, 3, 4, and 7), we found that the MW peaks are delayed relative to the corresponding HXR peaks by 1.9 to 3.0 s, with uncertainties less than 1 s. We interpret this result in light of the results on the spatial analysis we conducted in previous sections. In previous sections, we showed that, in both HXR and MW, there were two different emission sources. Since HXR and MW source locations were observed to be close to each other, we concluded that there were two different magnetic reconnection sites responsible for emission in both HXR and MW. Although peak 1 shows MW and HXR sources that are not spatially coincident (Figure 5), their close similarity of time profile suggests that MW and HXR emissions are closely connected, so we include it in the analysis. Thus, we compared delays among peaks 1, 3, 4, and 7. As seen in Table 1, since each delay is within the error bars of the others, it is hard to draw conclusions about any correlations between delays and other parameters such as MW energies, the peak X-ray or MW flux, or their ratio. Note, however, that at 9.4 GHz, the delay for peaks in the second source (peaks 3, 4 and 7) seems to increase toward later peaks. This corresponds to the general flux increase toward later peaks, seen most prominently at 9.4 GHz, and also seen in 17 GHz and 35 GHz (see Figure 2: peak 7 is the strongest peak observed at $\sim$ 01:55). This correlation is explicable in terms of the idea introduced by Gary \& Tang (1985), that the delay time of MW respect to HXR can be a result of the longer trapping time of MW producing electrons compared to that of HXR signal. To test this, we calculated the decay times of the 17 GHz MW signal at peak 3, 4, and 7 (by fitting each curve to simple exponentially decay function for few seconds after the peak) to be 6.05$\pm$1.29, 8.49$\pm$1.22, and 27.5$\pm$3.55 s, respectively, which increases toward later peaks, but the decay time of peak 7 is clearly strongly influenced by blended temporal peaks, and so is unreliable.  To summarize, our investigation of delay times shows MW delayed with respect to HXR, but relative delay and decay times do not show a clear pattern in this event, due primarily to the breadth of the peaks, which produces relatively large uncertainties.

\section{Discussion and Conclusion}

The comprehensive observations of the 2011 February 15 X2.2 flare provide an unique opportunity to investigate the properties of accelerated flare electrons. Besides coordinated coverage in both MW and HXR imaging spectroscopy that are not often available, this event has multiple emission peaks, allowing us to investigate possible different magnetic reconnection and particle acceleration processes associated with the individual peaks. We obtained two major results in this study.

(1)	We investigated  time delays of MW peaks in respect to
HXR peaks.  Although this delay was found in several previous studies (e.g. Cornell et al., 1982,  Gary \& Tang, 1985; Silva et al. 2000), our study allows a closer examination of such delay in the multiple peaks of a single event. We found that the MW lags HXR (50 -- 75 keV) by 1.9 to 3.0 seconds, with an uncertainty below 1 second. We also found that this delay is not related to other parameters:  the delay is similar for different MW frequencies, such as in 9.4 and 17 GHz; the delay does not depend on the peak flux or power index of the HXR and MW emissions; there is no difference in the delay for the two flare stages that will be discussed in our second result below.  Our study does not provide additional information beyond the existing explanation of the delays such as the concept of “trap-plus-precipitation”, and differences in the energy range of electrons that produces MW and HXR, as we discussed in the Introductio n (Silva et al. 2000).

(2) Applying image de-convolution techniques, we identified source locations of each peak in both HXR and MW. HXR and MW basically evolve with similar patterns. We found two distinct phases of the emissions,  the first stage with one emission peak, and the second stage containing 5 remaining peaks. The source morphology and location change substantially between these two stages. Therefore, we postulate that, in this flare, there are two distinct magnetic reconnection sites that produced HXR and MW emissions. For HXR, we found that there is a sharp transition in a time duration as short as 30 seconds between the two stages. For MW, it is interesting to note that the first source does not immediately disappear. In fact, it remains at a site of MW emission until the last, strongest peak starts. This may be related to the result (1) concerning the trapping of electron that delays MW emissions.

It is important to understand the physical mechanisms of the two stages of emission. It is clear that they are associated with energy release in two different sites. Within each stage, the sub-peaks seem to be at the same sites. Therefore, there must be two different groups of magnetic fields that are reconnected at the two stages. Wang et al. (2102) showed that the development of this flare can be explained by the tether cutting scenario originally proposed by Moore et al. (2001). Inoue et al. (2014) has performed a data-driven MHD simulation for this specific event, and they identified two clear stages of magnetic reconnection, consistent with the tether cutting scenario as well. Our observation also seems to support the tether-cutting scenario, perhaps with with some variation, as explained in the following. In the tether cutting scenario, a flare occurs in two stages, starting from the quadrupolar loop configuration. In the first stage, two J-shaped loops reconnect to form a set of two new loops, one small loop falling down toward the surface and another large loop erupting away from the surface. In the second stage, the erupting large loop cuts through the arcade fields as it rises, reconnecting as it meets new, larger loops. Wang et al. (2012) observed four HXR footpoints that can create two J-shaped loops, and their conjugate ends correspond to the peak 3 HXR double-footpoint sources we observed. They observed an increase in horizontal magnetic field strength in the area between these double-footpoint sources, which was interpreted as the evidence for the first stage of the tether cutting (a collapsing small loop was formed between these double-footpoint, increasing the horizontal magnetic field strength). Our observation seems to indicate that this "first stage" occurred at peak 2 as well. The schematic pictures explaining our version of the tether cutting scenario are shown in Figure 12. In Figure 12, the red circles indicate the HXR double-footpoint sources observed at peak 2 and the green circles indicate the HXR double-footpoint sources observed at peak 3. The yellow circles indicate the two remote footpoint sources that were observed by Wang et al. (2012) (we observed one of them as well, see Figure 3 or 5). The purple lines indicate the suggested flaring loops. The AIA 94 image taken at pre-flare time, 01:47:14 UT, is shown in (d) for comparison.  The blue stars in (a) and (b) indicate the two different reconnection sites. We propose the following scenario for the development of this flare. First, a reconnection occurs between loop 1 and loop 2, leaving a small loop between red circles and a large loop between northern green circle and western yellow circle. This corresponds to peak 2, where HXR is emitted from the locations corresponding to two red circles and western yellow circle, and MW is emitted from the small loop formed between two red circles. Next, a second reconnection occurs between the large loop formed in the first reconnectoin and loop 3. This leaves a second small loop between two green circles and also a second large loop between two yellow circles. This corresponds to peak 3, where HXR is emitted from the locations corresponding to two green circles and western yellow circle, and MW is emitted from the second small loop between two green circles. After peak 3, MW emission continues from the second small loop (and the first small loop). For HXR, the second large loop formed in peak 3 rises and cuts through the arcade fields, leaving the footpoint emissions through peak 4 to 7. This interpretation is based on the fact that we see the eastern HXR double-footpoint sources move around - along and away from the flare ribbon - as the flare progresses beyond peak 3 (see Figure 5). On the other hands, MW sources after peak 3 show almost no movements, which suggests to us that the emission was coming from relatively stable loop. Our scenario basically follows what was already proposed by Wang et al. (2012), but has a slightly new detail because of the additional sources that were observed simulteneously in HXR and MW at peak 2, \textit{before} the first stage claimed by Wang et al. (2012). It suggests to us that, at least for this flare, the first stage of the tether cutting may have been temporally and spatially distributed.

We also estimated the sizes of flaring loops based on the HXR and MW observation. From HXR observation, the height of the loop at peak 3 is about 4,300 km (semi-circular loop shape assumption). For the same peak, MW emission is generated at the height of approximately 10,000 to 20,000 km (assuming that the projection effect is displacing the true source location 5 -- 10 arcseconds). According to the simulation results by Inoue et al. (2014), the height of the top of the ascending loop at the time equivalent to right after the peak 3 in our study is approximately 21,600 km. These approximation roughly fits to our proposed scenario, that (at peak 3) the small loop left behind after the first stages of the tether cutting became the source of HXR emission at its footpoints, and the large loop that was newly formed started to erupt. Meanwhile, MW was emitted from the height between the small loop and the erupting loop - possibly from the top of the small loop.

Finally, our observations address the asymmetry between MW and HXR sources that was discussed by Sakao (1994) and Wang et al. (1995). Those authors found that high-frequency MW sources tend to be located in strong magnetic field regions, while HXR are weaker there due to reflection of precipitating electrons by converging fields. Instead, HXR sources are stronger at weaker field regions, where MW are not as efficiently produced at high frequencies. Analyzing the magnetic structure of the AR, we found that all MW sources exhibit a gradient in degree of polarization, with the greater degree of polarization on the negative polarity side of the source.  Since the NoRH maps show the sources to be located entirely on the positive-polarity side of the PIL, we believe that the true location of the radio sources should be 5-10$\arcsec$ north of where they appear, but are shifted southward due to the uncertainty in NoRH position calibration and/or by the projection effects. Since we found the stronger HXR sources to be on the positive-polarity side of the PIL, shifting the MW sources north would also result in consistency with the Sakao (1994) and Wang et al. (1995) results.  Our study would further motivate the modeling in two areas: the modeling of both HXR and MW emission based on the input accelerated electron distribution, and the modeling of the acceleration process of electrons during the magnetic reconnection.

\acknowledgements
This work was supported in part by NSF grants AST-1312802, AGS-1348513, AGS-1408703, and NASA grants NNX14AC87G and NNX13AG13G to New Jersey Institute of Technology. The authors are grateful to Dr. S. White for providing us with the brightness temperature data of NoRP 17 GHz channel, and Dr. K. Iwai for helping us obtain NoRH and NoRP data. We thank SDO and RHESSI teams for the data used in this publication.

\begin{deluxetable}{cccccccc}
\tabletypesize{\scriptsize}
\tablewidth{0pt}
\rotate
\tablecaption{HXR peak times and MW delay times (and their uncertainties). This unit is second.}
\tablehead{\colhead{\bf Peak number}  & \colhead{\bf 1} & \colhead{\bf 2} & \colhead{\bf 3} & \colhead{\bf 4} & \colhead{\bf 5} & \colhead{\bf 6} & \colhead{\bf 7}}
\startdata
{\bf HXR 50 -- 75 keV} & 01:49:22.3 (0.127)  & 01:52:41.1 (0.210)  & 01:53:11.5 (0.177) & 01:53:39.9 (0.357) & 01:54:00.3 (0.196) &  01:54:08.3 (0.638) & 01:55:15.9 (0.247) \\
{\bf MW 9.4 GHz delay time} & 2.4 (0.48) & N/A (N/A) & 1.9 (0.80) & 2.1 (0.72) & 0.6 (0.55) & N/A (N/A) & 3.0 (0.73) \\
{\bf MW 17 GHz delay time} & 2.5 (0.53) & 4.8 (0.25) & 2.1 (0.90) & 1.6 (0.69) & 0.9 (0.53) & 0.7 (0.91) & 2.0 (0.66)
\enddata
\end{deluxetable}

\begin{figure}
\centering
\epsscale{1.0}
\plotone{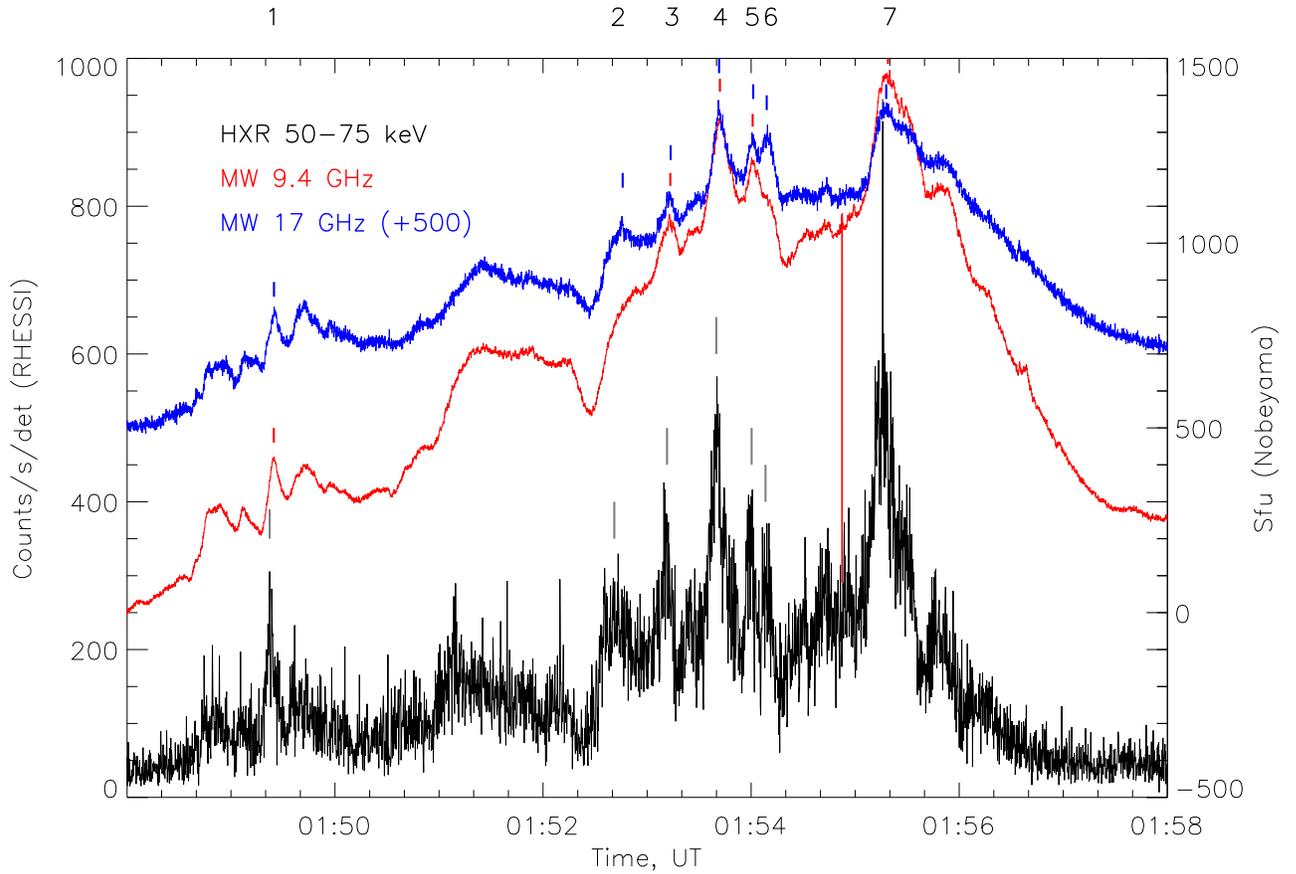}
\caption{HXR 50 -- 75 keV, MW 9.4 GHz, and MW 17 GHz time profile from 01:48:00 to 01:58:00. HXR cadence is 0.2 to 0.3 second and MW cadence is 0.1 second. Grey, red, and blue short lines indicate the HXR, MW 9.4 GHz, and MW 17 GHz peaks, respectively, which we identified and calculated the peak times. Note that we only chose combinations of peaks that were distinct enough to calculate the peak times, and MW 17 GHz flux is artificially increased by 500 SFU in order for it to be on the same plotting window as 9.4 GHz curve. \label{f1}}
\end{figure}

\begin{figure}
\centering
\epsscale{1.0}
\plotone{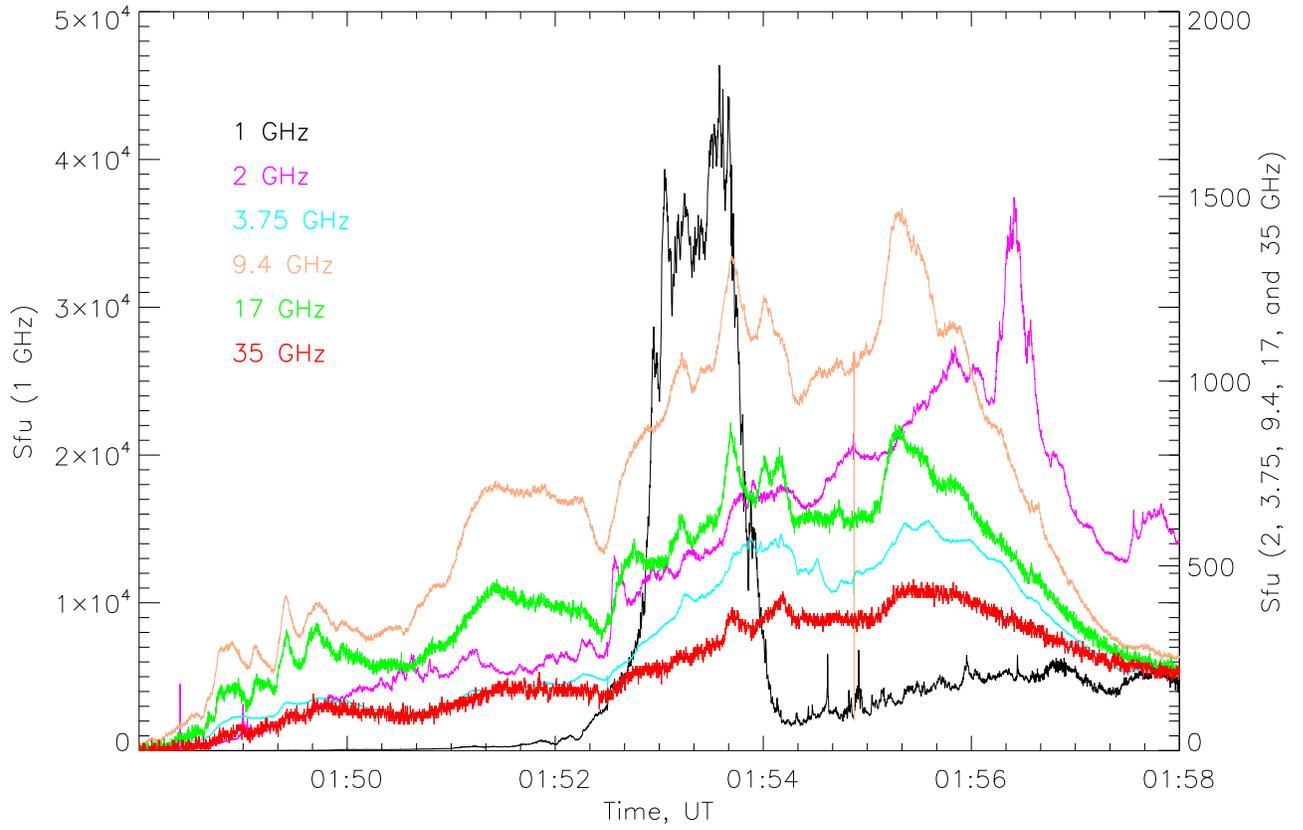}
\caption{NoRP time profiles from 01:48:00 to 01:58:00 in 1, 2, 3.75, 9.4, 17, and 35 GHz frequencies. All are backgroud subtracted. Note that the Y-axis' scale for 1 GHz is different from the Y-axis' scale for 2, 3.75, 9.4, 17, and 35 GHz. \label{MWlightcurve}}
\end{figure}

\begin{figure}
\centering
\plotfiddle{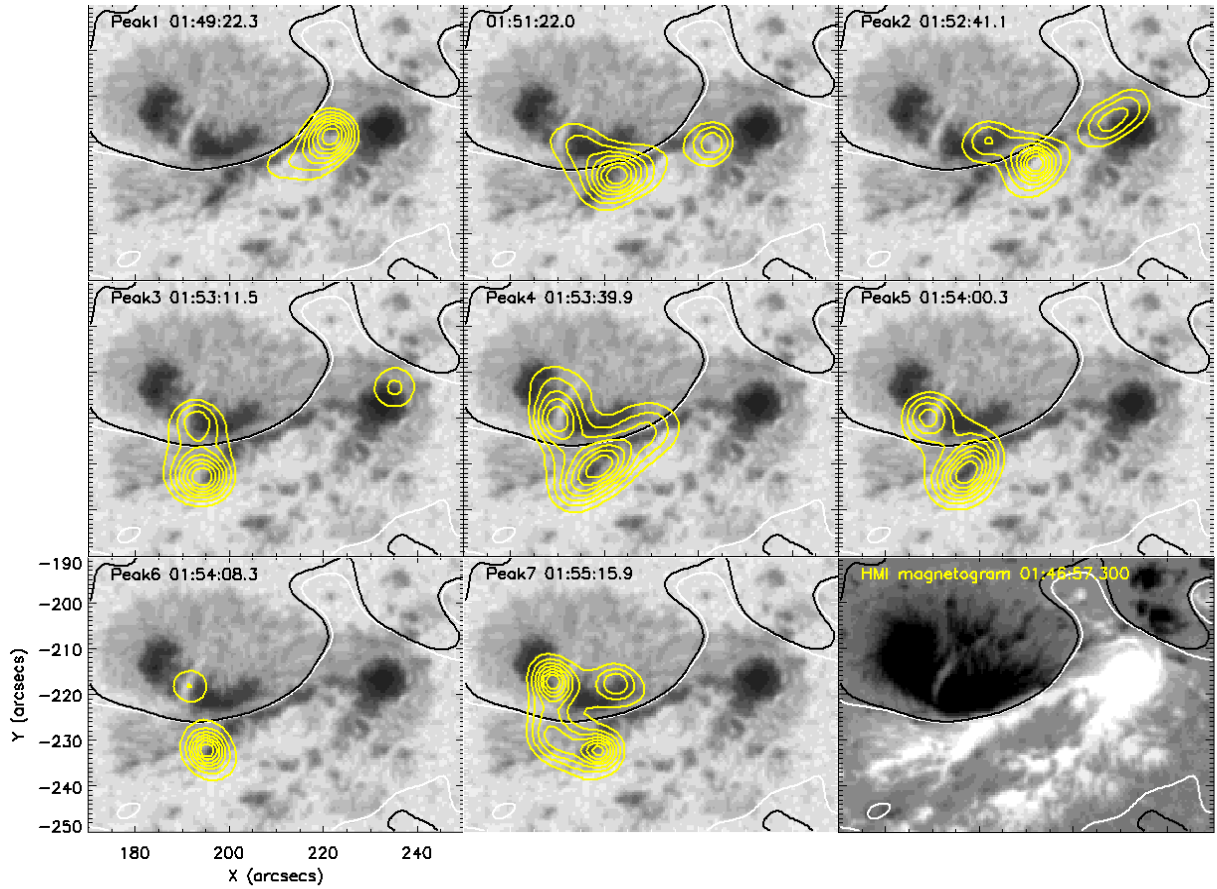}{1in}{0.}{400.}{280.}{-70}{0}
\caption{RHESSI PIXON image intensity contours in 50 -- 75 keV for peaks 1-7. One frame is additionally shown between peak 1 and 2, at 01:51:22 UT (40 seconds integration) to show the flare development during this long interval (see Figure 1). Contour levels are 30 -- 90 percent of the maximum. White and black contours indicate the positive and negative sides of the polarity inversion lines, respectively.}
\end{figure}

\begin{figure}
\centering
\plotfiddle{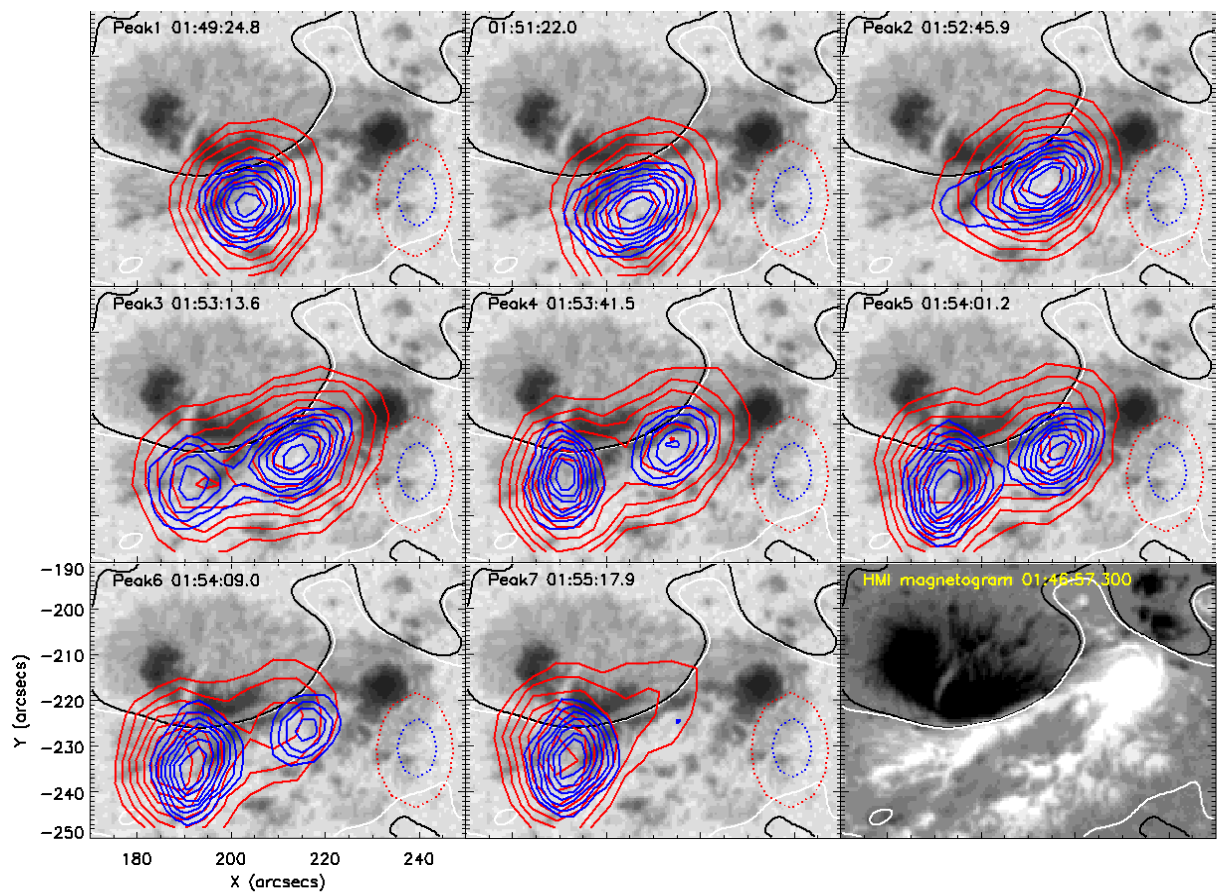}{1in}{0.}{400.}{280.}{-70}{0}
\caption{NoRH images' intensity contours in 17 GHz (red) and 34 GHz (blue) at peak times for peak 1-7 in 17 GHz time profile shown in Figure 1. One frame is additionally shown between peak 1 and 2, at 01:51:22 UT to show the flare development during this long interval (see Figure 1). Contour levels are 30 -- 90 percent of the maximum. The dotted red and blue circles show the beam sizes (contour of half power beam width) at each time for 17 GHz and 34 GHz, respectively. White and black contours indicate the magnetic polarity inversion lines, as in Figure 3.}
\end{figure}

\begin{figure}
\centering
\plotfiddle{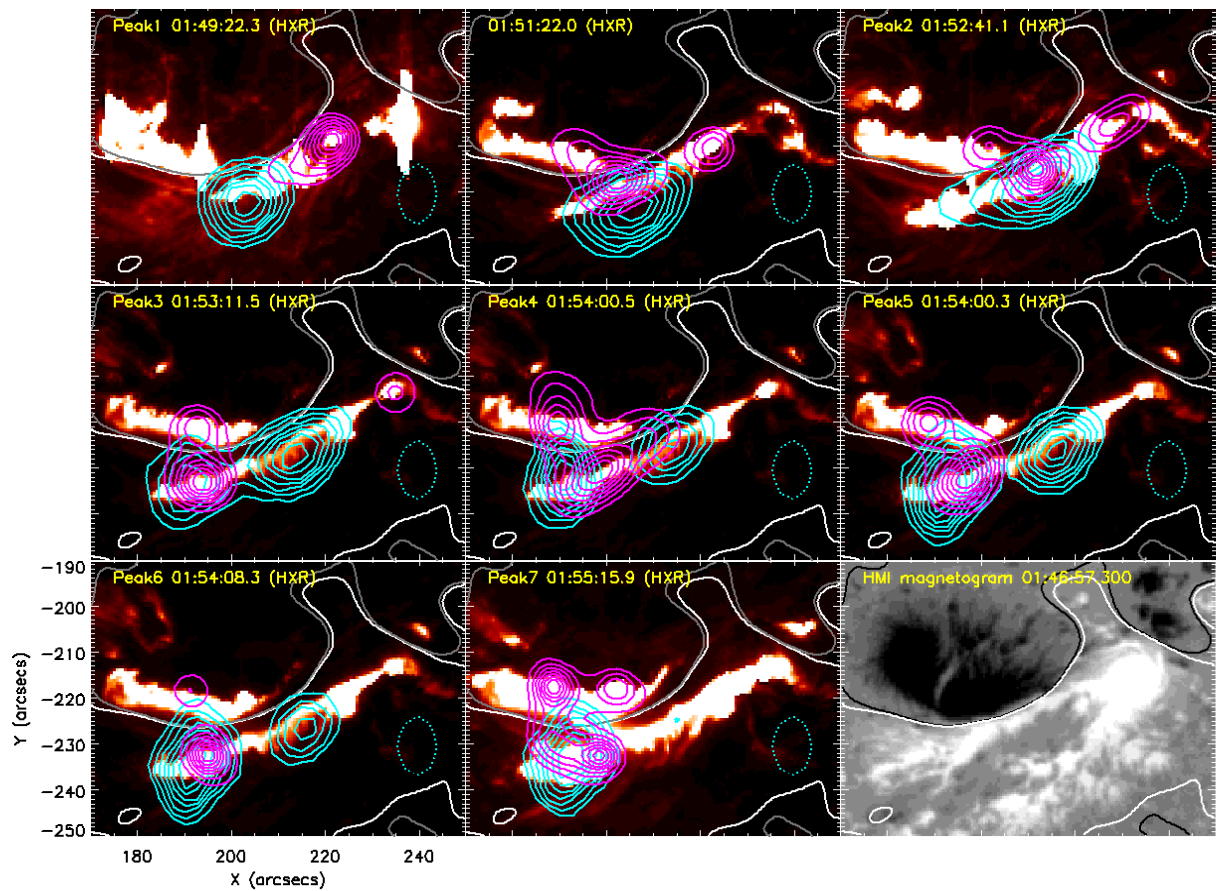}{1in}{0.}{400.}{280.}{-70}{0}
\caption{MW 34 GHz intensity contours (cyan) overlapped with HXR 50 -- 75 keV energy contours (magenta) on top of AIA HeII 304\AA images with the polarity inversion line, for peak 1-7. The dotted cyan circles show the beam size (contour of half power beam width) at each time for 34 GHz. One frame is additionally shown between peak 1 and 2, at 01:51:22 UT to show the flare development during this long interval (see Figure 1).}
\end{figure}

\begin{figure}
\centering
\plotfiddle{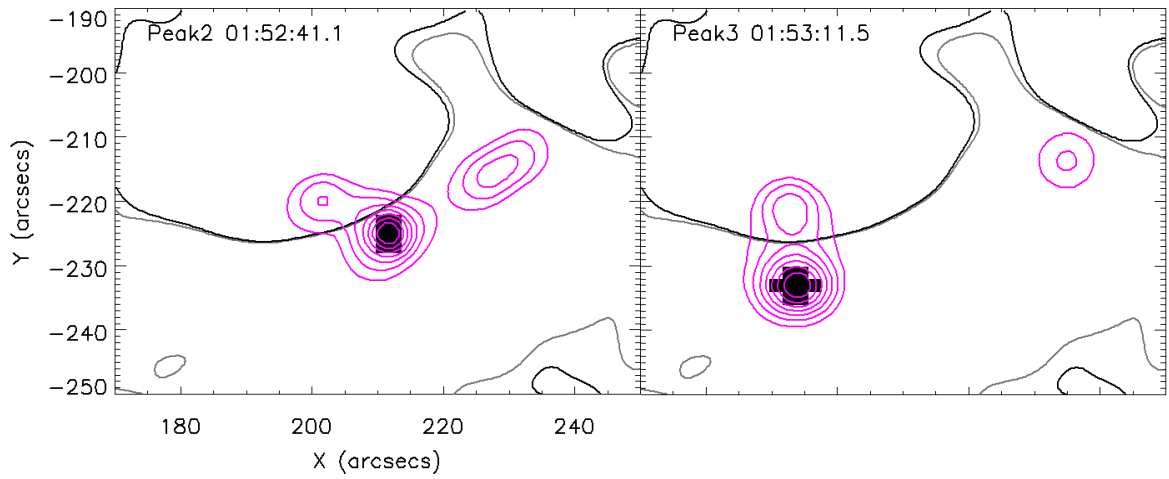}{0in}{0.}{400.}{280.}{-70}{0}
\caption{Identified HXR kernel pixels at peak 2 and 3, with threshold of 80 percent of the maximum intensity of the entire image.}
\end{figure}

\begin{figure}
\centering
\epsscale{1.0}
\plotone{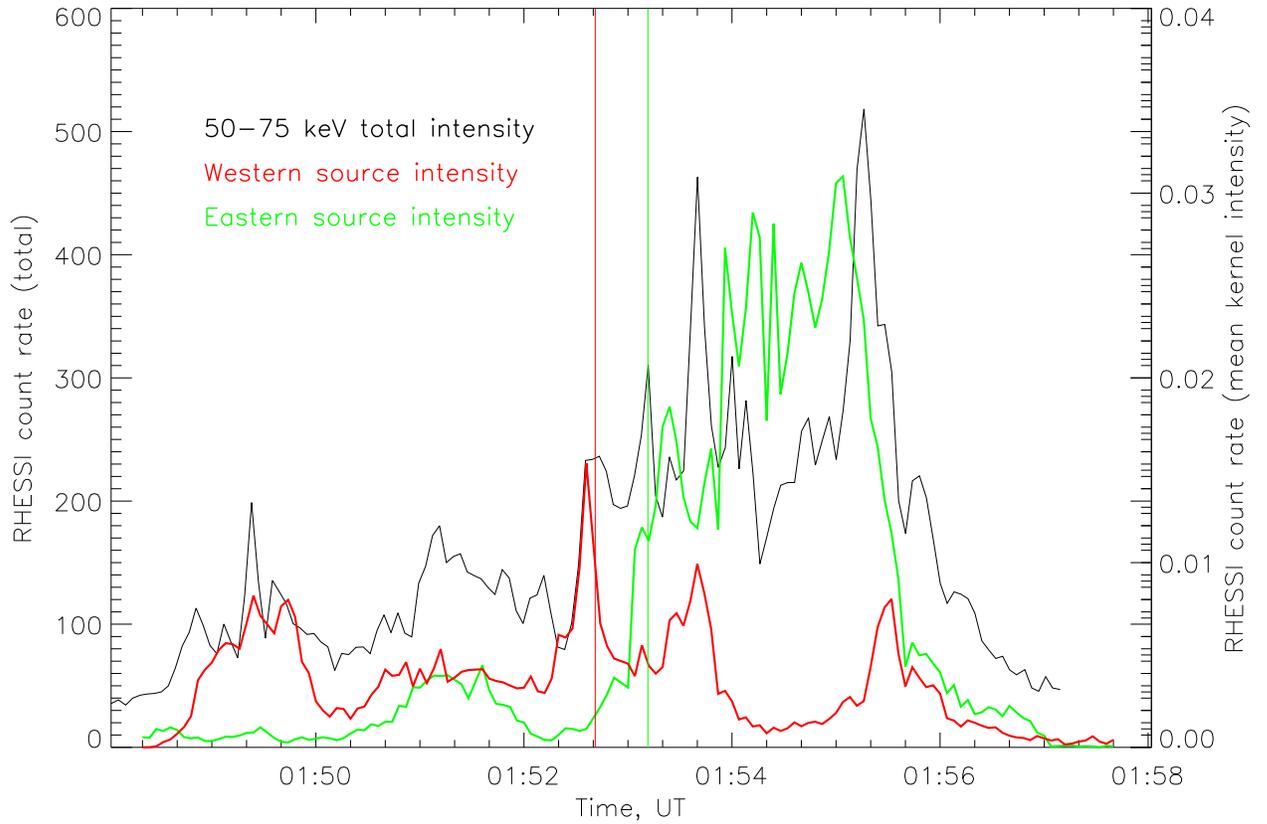}
\caption{HXR total intensity vs. mean kernel intensity time profile. The black curve is the original HXR lightcurve that was shown solely in Figure 1, presented with the same resolution as other two colored curves (4 seconds). The red curve is the time profile for the western source, and the green curve is the time profile for the eastern source. The red vertical line marks 01:52:41.1, the peak time at peak 2, and the green vertical line marks 01:53:11.5, the peak time at peak 3.}
\end{figure}

\begin{figure}
\centering
\plotfiddle{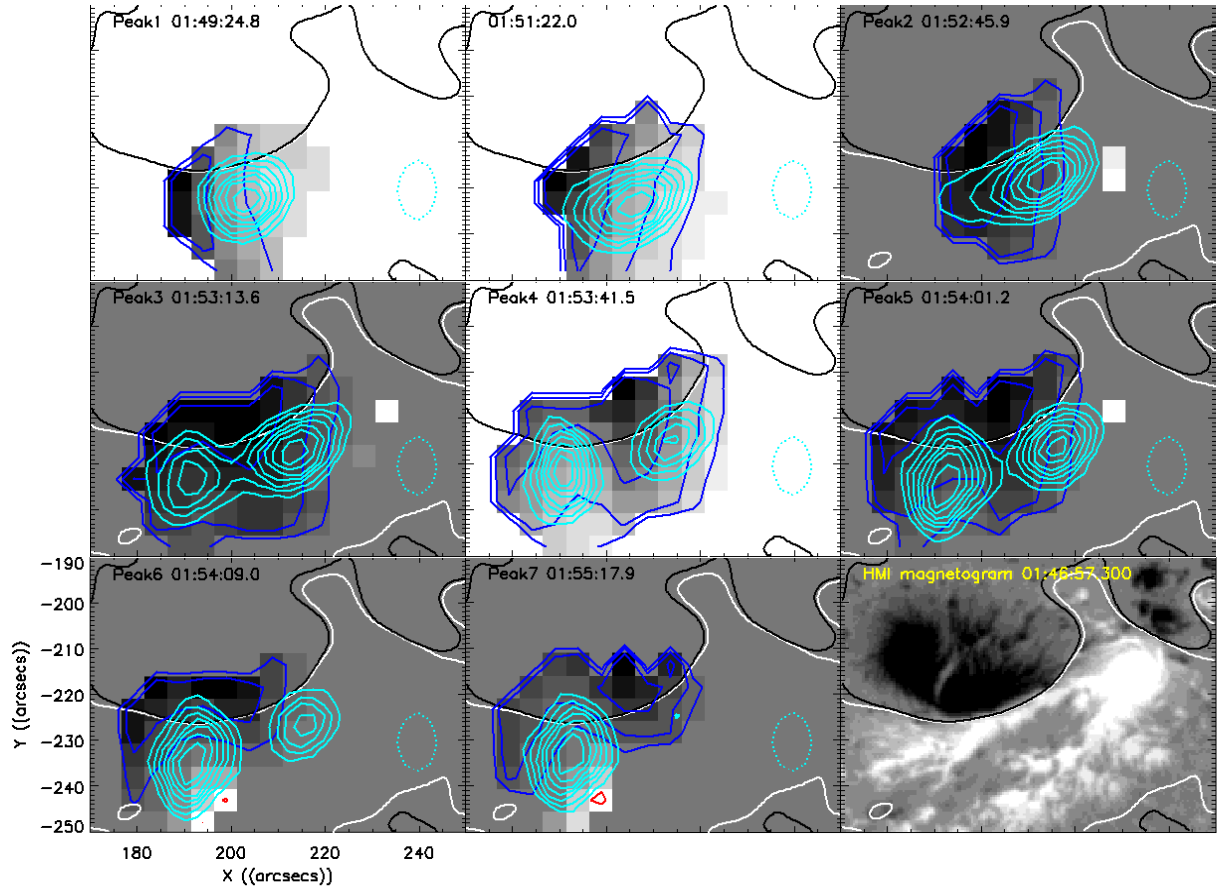}{1in}{0.}{400.}{280.}{-70}{0}
\caption{NoRH degree of polarization map (17 GHz) for peak 1 -- 7. Red contours are the degrees of right-circular polarization and blue contours are the degrees of left-circular polarization, both in the scale of 5, 10, 20, 40, 60, and 80 percent. MW 34 GHz intensity contours are overlapped in cyan, with the dotted circles showing the beam size (contour of half power beam width) at each time. One frame is additionally shown between peak 1 and 2, at 01:51:22 UT to show the flare development during this long interval (see Figure 1).}
\end{figure}

\begin{figure}
\centering
\plotfiddle{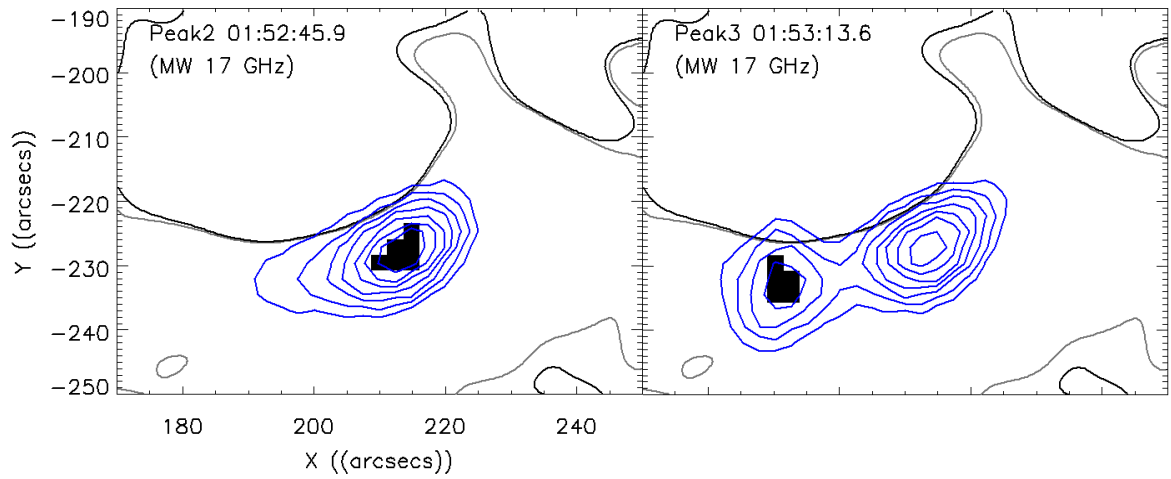}{0in}{0.}{400.}{280.}{-70}{0}
\caption{Identified MW kernel pixels at peak 2, with threshold of 90 percent of the maximum intensity of the entire image, and at peak 3, with threshold of 60 percent of the maximum intensity of the entire image.}
\end{figure}

\begin{figure}
\centering
\epsscale{1.0}
\plotone{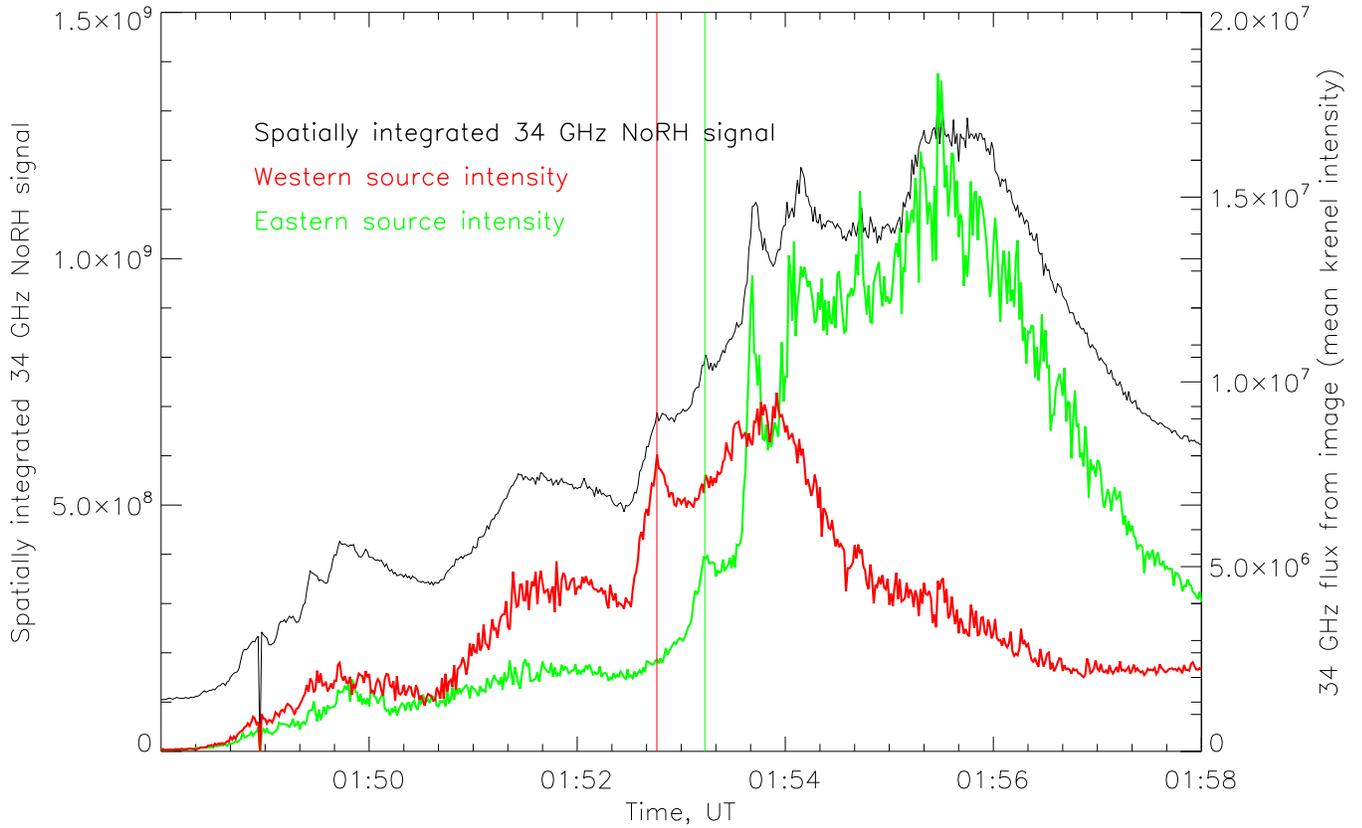}
\caption{Spatially integrated 34 GHz NoRH signal vs. mean kernel intensity time profile. The black curve is the spatially integrated signal from 34 GHz NoRH images. The red curve is the time profile for the western source, and the green curve is the time profile for the eastern source. The red vertical line marks 01:52:45.9, the peak time at MW 17 GHz peak 2, and the green vertical line marks 01:53:13.6, the time of peak 3 at MW 17 GHz peak 3.}
\end{figure}

\begin{figure}
\centering
\epsscale{1.0}
\plotone{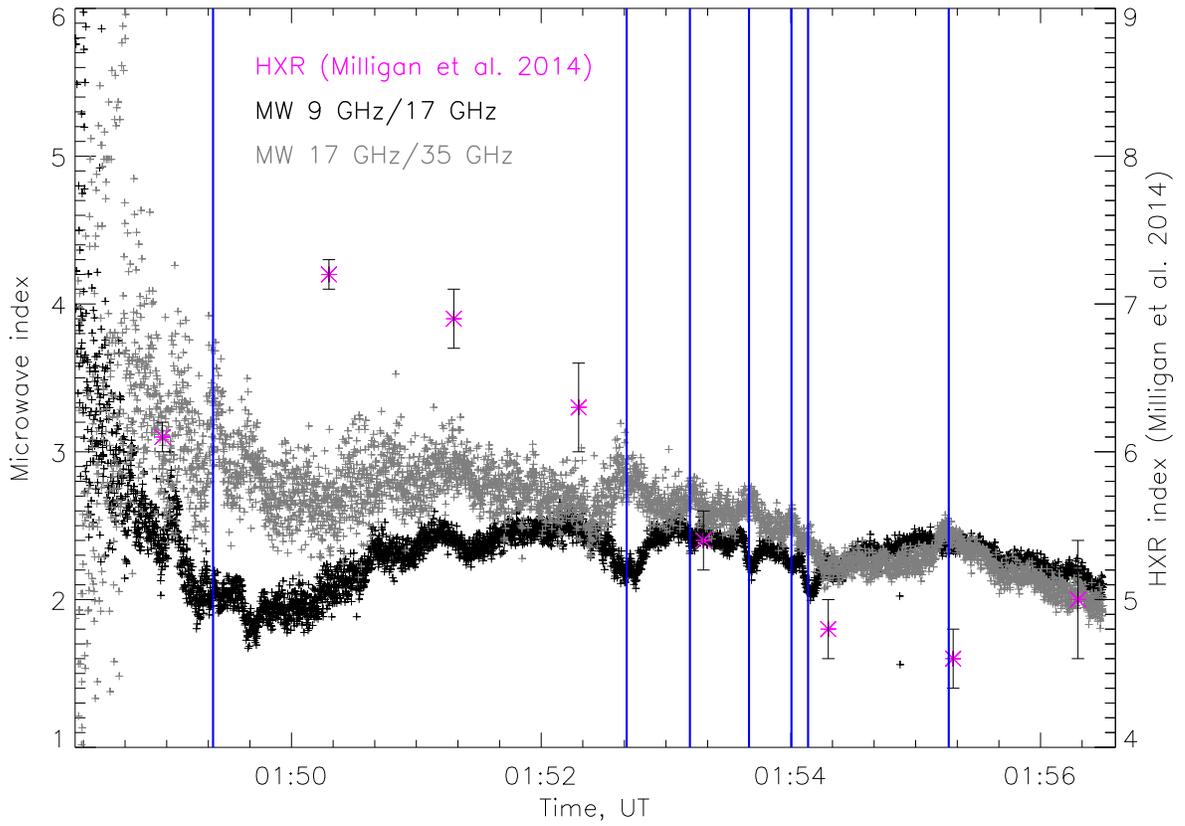}
\caption{Time profile of electron spectral indices inferred from the MW observations. The electron power indices inferred from the HXR observations, calculated by Milligan et al. (2014), are also plotted. Blue vertical lines indicate time of identified HXR peaks. }
\end{figure}

\begin{figure}
\centering
\epsscale{1.0}
\plotone{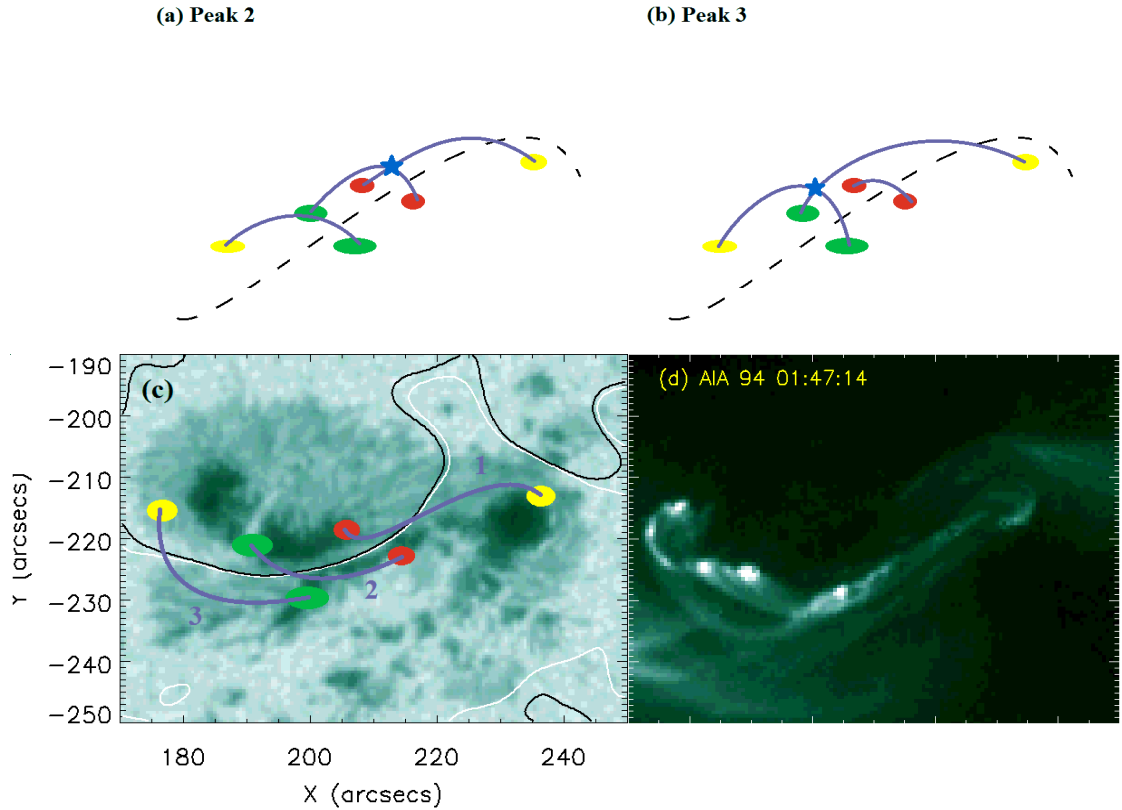}
\caption{The schematic pictures demonstrating the suggested geometry of the flaring loops based on the analysis of HXR and MW sources. (a) and (b): The scenarios for peak 2 and 3, respectively. The red circles correspond to the western HXR double-footpoint sources at peak 2, the green circles correspond to the eastern HXR double-footpoint sources at peak 3, the purple lines correspond to the suggested flaring loops, and the yellow circles correspond to two of the footpoint sources that were observed by Wang et al. (2012), which were expected in their quadrupolar loop configuration. The blue stars indicate possible reconnection sites. (c): Top view of the suggested loop configuration. (d): AIA 94 image taken at pre-flare time, 01:47:14 UT. See texts for details.}
\end{figure}

\end{document}